\documentclass[fleqn,10pt]{wlscirep}

\title{Architecture and Function of Mechanosensitive Membrane Protein Lattices}

\author[1,*]{Osman Kahraman}
\author[2]{Peter D. Koch}
\author[3]{William S. Klug}
\author[1,+]{Christoph A. Haselwandter}
\affil[1]{Departments of Physics \& Astronomy and Biological Sciences, University of Southern California, Los Angeles, CA 90089, USA}
\affil[2]{Department of Systems Biology, Harvard Medical School, Boston, MA 02115, USA}
\affil[3]{Department of Mechanical and Aerospace Engineering, University of California, Los Angeles, CA 90095, USA}

\affil[*]{osman.kahraman@usc.edu}
\affil[+]{cah77@usc.edu}

\begin{abstract}
Experiments have revealed that membrane proteins can form two-dimensional clusters with regular translational and orientational protein arrangements, which may allow cells to modulate protein function. However, the physical mechanisms yielding supramolecular organization and collective function of membrane proteins remain largely unknown. Here we show that bilayer-mediated elastic interactions between membrane proteins can yield regular and distinctive lattice architectures of protein clusters, and may provide a link between lattice architecture and lattice function. Using the mechanosensitive channel of large conductance (MscL) as a model system, we obtain relations between the shape of MscL and the supramolecular architecture of MscL lattices. We predict that the tetrameric and pentameric MscL symmetries observed in previous structural studies yield distinct lattice architectures of MscL clusters and that, in turn, these distinct MscL lattice architectures yield distinct lattice activation barriers. Our results suggest general physical mechanisms linking protein symmetry, the lattice architecture of membrane protein clusters, and the collective function of membrane protein lattices.
\end{abstract}
\begin{document}

\flushbottom
\maketitle

\thispagestyle{empty}

\section*{Introduction}
Superresolution light microscopy and electron cryo-tomog\-raphy have revealed \cite{Baddeley2009,Briegel2009,Greenfield2009,Specht2013}
that integral membrane proteins can form large clusters with regular and distinctive translational and orientational protein arrangements. 
Cooperative interactions in such membrane protein lattices may provide a general mechanism for cells to modulate protein function \cite{Bray2004,Engelman2005}. Self-assembly of membrane protein lattices requires energetically favorable direct protein-protein \cite{Park2006,Sieber2007,Briegel2014b} or indirect lipid bilayer-mediated interactions \cite{Harroun99,Goforth2003,Botelho2006} and, for the ground-state architecture of planar lattices to be anything other than hexagonal, interactions must be directional. Directionality of
bilayer-mediated interactions can be induced by the discrete symmetry of membrane proteins, which occur in a variety of different oligomeric states \cite{Linden2012,Gandhi2011,Walton2015}. Molecular dynamics simulations have suggested \cite{Periole2007,Parton2011,Mondal2013,Yoo2013} that bilayer-mediated interactions can yield ordering of membrane proteins. While the membrane
elasticity theory underlying bilayer-mediated protein clustering has been studied in some detail \cite{Phillips2009,Goulian1993,Weikl1998,Kim1998,Fournier1999,Kim2008,Frese2008,Auth2009,Muller2010,Reynwar2011,Bahrami2014,Evans2003,Dommersnes1999,Weitz2013,Yolcu2014,Golestanian1996,Weikl2001,Lin2011,Dan1993,ArandaEspinoza1996,harroun99b,Partenskii2004,Brannigan2007,Ursell2007,CAH2014a}, only little is known about the lattice architectures due to elastic interactions between specific integral membrane proteins, and how lattice architecture and elastic interactions affect protein~function.

In this Article we study the most favorable (minimum-energy) lattice architectures, and corresponding modulation of protein function, due to bilayer-mediated elastic interactions between mechanosensitive membrane proteins. A diverse range of integral membrane proteins have been shown to be mechanosensitive \cite{Phillips2009,Anishkin2014} and, in particular, the gating of prokaryotic~\cite{Kung2010} and eukaryotic~\cite{Brohawn2014} ion channels depends on the mechanical properties of the surrounding lipid bilayer. We employ the bacterial mechanosensitive channel of large conductance (MscL) \cite{Phillips2009,Kung2010} as a model system to develop relations between protein symmetry, lattice architecture, and the collective function
of membrane protein lattices. 

MscL switches from a closed to an open state with increasing membrane tension \cite{Phillips2009,Kung2010}. 
Protein crystallography has yielded tetrameric \cite{Liu2009} as well as pentameric \cite{Chang1998,Walton2013} MscL structures. The physiological significance of pentameric MscL is well established \cite{Sukharev1999,Dorwart2010}. In contrast, direct experimental evidence of tetrameric MscL has so far only been obtained \textit{in vitro} \cite{Dorwart2010,Iscla2011,Gandhi2011,Reading2015}, it is uncertain whether MscL can occur as a tetramer \textit{in vivo}, and the physiological significance of tetrameric MscL is a matter of debate \cite{Dorwart2010,Iscla2011,Gandhi2011,Reading2015,Walton2015}. In particular, it has been proposed that MscL can only occur as a pentamer \textit{in vivo} \cite{Dorwart2010,Iscla2011} or that, while pentamers are predominant, MscL can occur as a mixture of different oligomeric states \textit{in vivo} \cite{Gandhi2011,Walton2015}, with different MscL oligomeric states having specific functional roles or serving as assembly intermediates. Interestingly, a number of different experiments have shown \cite{Dorwart2010,Iscla2011,Gandhi2011,Reading2015} that the oligomeric state of MscL can be modified by selectively truncating MscL, tuning the lipid or detergent compositions used in \textit{in vitro} experiments, or varying the temperature. In this Article we take the available MscL structures as our starting point, and consider the lattice architectures and collective functions of clusters of both tetrameric and pentameric MscL, as well as mixtures of tetrameric and pentameric MscL.

\textit{In vitro} and \textit{in vivo} studies have suggested that bilayer-mediated interactions stabilize large clusters of hundreds of MscL \cite{Grage2011}, that MscL activation is affected by clustering \cite{Grage2011,Nomura2012}, and that MscL number is strongly regulated in response to environmental stimuli \cite{Bialecka2012}, indicating \cite{Grage2011,Nomura2012} that bacteria may use MscL clustering, and bilayer-mediated interactions, to modulate MscL function. In the remainder of this Article, we first describe how bilayer-mediated interactions can be efficiently calculated for the large MscL clusters observed in experiments, and then use this approach to predict the minimum-energy lattice architectures for tetrameric and pentameric MscL, and to suggest how differences in lattice architecture affect MscL activation.

\section*{Methods}
\subsection*{Bilayer-mediated protein interactions}
\begin{figure}
\center
 \includegraphics[width=0.9\textwidth]{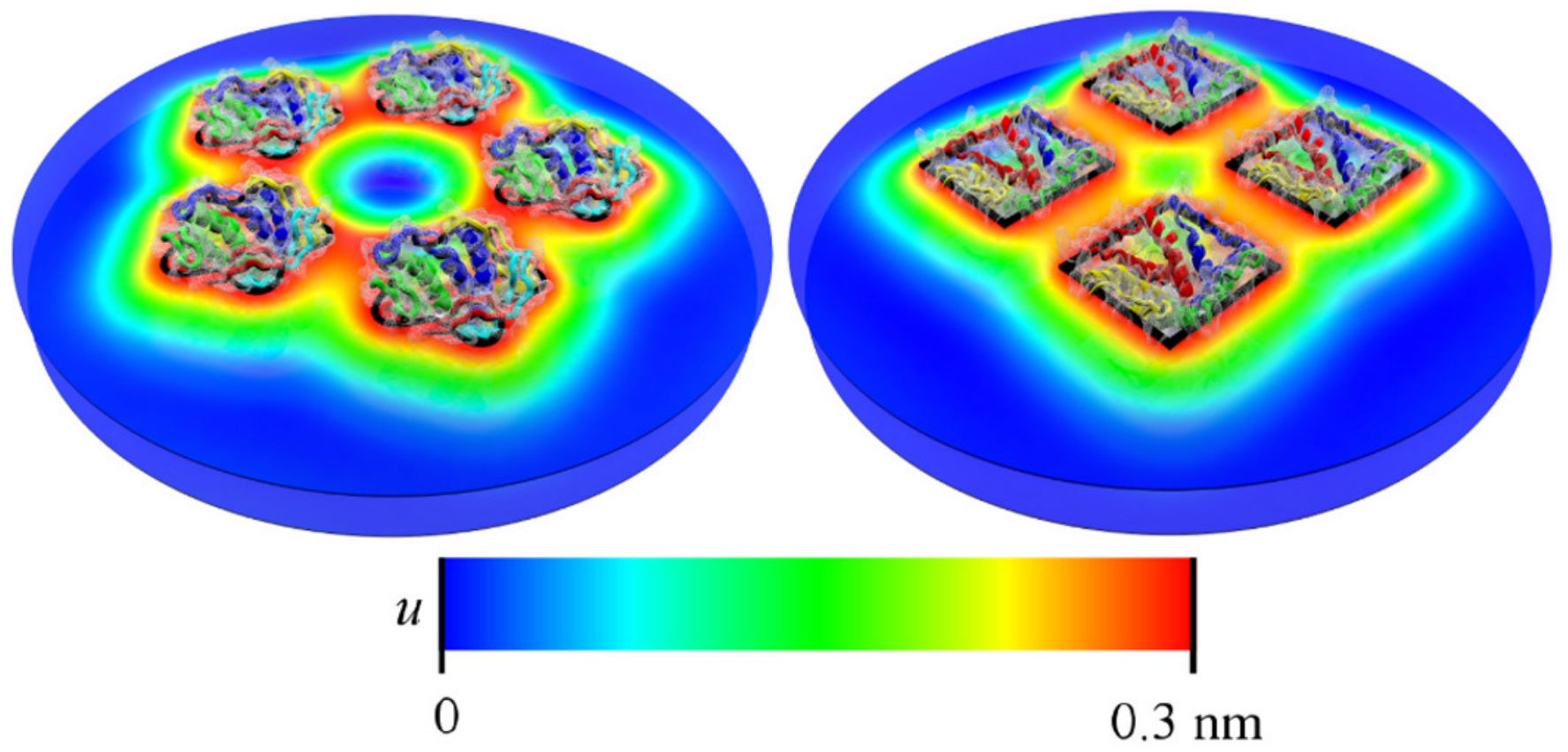}
\caption{\label{fig:surfaces} 
\textbf{Overlapping bilayer thickness deformation fields induce thickness-mediated interactions between MscL.}
Pentameric (Protein Data Bank accession number 2OAR) \cite{Chang1998} (left panel) and tetrameric (Protein Data Bank accession number 3HZQ) \cite{Liu2009} (right panel)
MscL structures, their five-fold clover-leaf and tetragonal representations \cite{CAH2013b} (black curves superimposed on MscL structures), and the corresponding MscL-induced thickness deformations $u$ calculated from equation~(\ref{eq:elastic_energy}) using our finite element approach for the indicated arrangements of closed MscL (see Fig.~\ref{fig:pair} for the thickness-mediated interaction energies associated with the MscL arrangements shown).
The MscL-induced bilayer thickness deformations depend on MscL shape, separation, and orientation, as well as on the effective bilayer properties captured by equation~(\ref{eq:elastic_energy}).
}
\end{figure}

Bilayer-mediated protein clustering may be driven by curvature deformations \cite{Goulian1993,Weikl1998,Kim1998,Kim2008,Auth2009,Muller2010,Frese2008,Reynwar2011,Bahrami2014,Fournier1999,Dommersnes1999,Evans2003,Weitz2013,Yolcu2014},
bilayer fluctuations \cite{Dommersnes1999,Evans2003,Weitz2013,Yolcu2014,Golestanian1996,Weikl2001,Lin2011},
or thickness deformations \cite{Dan1993,ArandaEspinoza1996,harroun99b,Fournier1999,Partenskii2004,Brannigan2007,Ursell2007,CAH2014a}.
Experiments and previous theoretical work on MscL suggest \cite{Ursell2007,Phillips2009,Grage2011,Nomura2012} that, at the small protein separations relevant for MscL clusters, thickness-mediated interactions between MscL are dominant (see Fig.~\ref{fig:surfaces}). We therefore focus on thickness-mediated interactions which, in the simplest formulation, are governed by an elastic energy of the form~\cite{AndersenKoeppe}
\begin{equation} \label{eq:elastic_energy}
\textstyle G=\frac{1}{2}\int \text{d}x \text{d}y \left\{K_b (\nabla^2 u)^2+K_t \left(\frac{u}{a}\right)^2+\tau
\left[2 \frac{u}{a}+(\nabla u)^2 \right] \right\} \,,
\end{equation}
where the thickness deformation field $u(x,y)$ is one-half the bilayer hydrophobic thickness mismatch, $K_b$ is the bending rigidity, $K_t$ is the thickness deformation modulus, $a$ is one-half the hydrophobic thickness of the unperturbed lipid bilayer, and, for generality, we consider \cite{OWC1} the coupling of the membrane tension $\tau$ to $u$ as well as to area changes. Experiments roughly yield $K_b = 20$ $k_B T$ and $K_t = 60$ $k_B T/$nm$^2$\cite{AndersenKoeppe,Phillips2009},
which we used here, but the values of these effective parameters \cite{Phillips2009} generally change with bilayer composition \cite{Rawicz2000}. 
Unless indicated otherwise, we set $\tau=0$. The continuum theory exemplified by equation~(\ref{eq:elastic_energy}) does not capture detailed molecular effects \cite{Parton2011,Periole2007,Mondal2013,Yoo2013,West2009,Kim2012}, but encapsulates many crucial properties of protein-induced bilayer deformations  \cite{Phillips2009,Goulian1993,Weikl1998,Kim1998,Frese2008,Kim2008,Auth2009,Muller2010,Reynwar2011,Bahrami2014,Dommersnes1999,Evans2003,Weitz2013,Yolcu2014,Golestanian1996,Weikl2001,Lin2011,Dan1993,ArandaEspinoza1996,harroun99b,Fournier1999,Partenskii2004,Brannigan2007,Ursell2007,CAH2014a,AndersenKoeppe}
and, in particular, has been found previously\cite{Grage2011,Nomura2012,Wiggins2004,Perozo2002,Phillips2009} to explain
key aspects of MscL clustering and gating.

\subsection*{Modeling tetrameric and pentameric MscL}
\label{secModel}
Based on structural data \cite{Chang1998,Liu2009}, we model tetrameric and pentameric MscL in the closed and open states as described in Ref.~65\nocite{CAH2013b}. We summarize here the quantitative details of these simple models of MscL shape. The molecular structure of MscL found in \textit{Mycobacterium tuberculosis} \cite{Chang1998} suggests a five-fold clover-leaf shape of pentameric MscL (see MscL in the left panel of Fig.~\ref{fig:surfaces}), which can be represented by the contour
\begin{align}
 C(\theta) = R\left[1+ \epsilon \cos(5(\theta-\omega)) \right]
\end{align}
in polar coordinates, where $R$ captures the size of MscL, $\epsilon$ is the amplitude of angular undulations, and $\omega$ denotes the orientation of MscL with respect to the $x$-axis. The observed structure of closed pentameric MscL\cite{Chang1998} suggests \cite{CAH2013b} an amplitude $\epsilon=0.22$ and radius $R=R^c_\text{penta}=2.27$~nm.
Based on proposed structures of MscL in the open state \cite{Sukharev2001a, Sukharev2001b}, we set $\epsilon=0.11$ and $R=R^o_\text{penta}=3.49$~nm for
open pentameric MscL. Similarly, we model the tetrameric structure of MscL found in \textit{Staphylococcus aureus} \cite{Liu2009} 
by tetragonal shapes in the closed and open states (see MscL in the right panel of Fig.~\ref{fig:surfaces}). To isolate the effects of MscL shape on bilayer-MscL interactions, we follow here Ref.~65 and use the same approximate areas for the transmembrane cross sections of tetrameric and pentameric MscL.

On the basis of structural data on MscL, the hydrophobic thickness of MscL in the closed and open states has been estimated \cite{Ursell2008} to be $h_c = 3.8$~nm and $h_o = 2.5$~nm, respectively. Thus, we use the boundary conditions $U_c = h_c/2 -a =0.3$~nm and $U_o = h_o/2 -a = -0.35$~nm for $u$ along the bilayer-protein interface in the closed and open states of MscL, where the unperturbed bilayer half-thickness $a=1.6$~nm approximately
corresponds to lipids extracted from \textit{E. coli} \cite{Mitra2004} as
well as other organisms \cite{AndersenKoeppe,Phillips2009}. Following
previous studies on MscL \cite{Wiggins2004,Wiggins2005,Ursell2007,Ursell2008},
we use zero-slope boundary conditions along the bilayer-protein interface.

\subsection*{Mixed finite element formulation}
\label{secFE}
While the anisotropic thickness deformations due to a few proteins can be obtained by minimizing equation~(\ref{eq:elastic_energy}) using perturbation analysis \cite{CAH2013a,CAH2013b} or finite-difference schemes~\cite{Ursell2007,Mondal2011}, calculation of the minimum-energy lattice architectures for large MscL clusters is not practical with either approach. The finite element method for solving boundary value problems yields rapid numerical convergence even for very complicated integration domains and, hence, provides a suitable approach for computing bilayer-mediated interactions in large protein clusters. However, standard finite element implementations are not able to account for the dependence of equation~(\ref{eq:elastic_energy}) on thickness stretch and gradient terms while satisfying the stringent continuity requirements necessitated by the curvature terms. To overcome this challenge we combine~\cite{OWC1} Lagrange shape functions for the thickness stretch and gradient terms with a discrete Kirchhoff triangle (DKT) formulation \cite{Batoz1980} for curvature deformations. 

Following the standard finite element discretization procedure, we rewrite the variation of the energy in equation~(\ref{eq:elastic_energy}) with respect to nodal 
degrees of freedom $\mathbf{U}$ as a summation over elements,
\begin{align}
 \delta G = \sum_{e\in \text{elements}}  \left(
 \delta \mathbf{U}^e)^T (\mathbf{K}^e \mathbf{U^e} +\mathbf{f}^e
 \right) \, ,
\end{align}
where the element stiffness matrix $\mathbf{K}^e$ and ``internal tension'' $\mathbf{f}^e$,
\begin{align}
 \mathbf{K}^e &= 2A^e \int \int \mathbf{B}^T \mathbf{D} \mathbf{B} \text{d}\xi\text{d}\eta \, , \\
 \mathbf{f}^e &= 2A^e \int \int \frac{\tau}{a}\mathbf{M}  \text{d}\xi\text{d}\eta \, ,
\end{align}
are integrated over the local coordinates $(\xi,\eta)$ of elements and weighted by the element areas $A^e$. 
The constitutive matrix $\mathbf{D}$ is a block diagonal matrix with the lipid bilayer parameters as coefficients.
The strain-displacement transformation matrix $\mathbf{B}$ combines the DKT shape functions $\mathbf{H}$ with  the linear triangular shape functions $\mathbf{M}$:
\begin{align}
 \mathbf{B}^T = 
  \left[
 \begin{array}{c}
  \mathbf{M}^T \qquad
  \mathbf{M}^T_{,x} \qquad
  \mathbf{M}^T_{,y} \qquad
  \mathbf{H}^T_{x,x} \qquad
  \mathbf{H}^T_{y,y} 
 \end{array}
\right]
\, .
\end{align}
Explicit forms of the DKT shape functions $\mathbf{H}$ are given by Batoz \textit{et al.}\cite{Batoz1980}, while the linear
triangular shape functions $\mathbf{M}$ can be found in standard finite element textbooks---see, e.g., Ref.~74\nocite{shames1985energy}.
We triangulated the bilayer surface using the ``Frontal'' algorithm from the Gmsh package\cite{TriangulateRef}. 
We implemented our finite element formulation in C++ using the variational mechanics library VOOM and minimized the energies using the L-BFGS-B solver \cite{Zhu1997}.
We checked for convergence using standard procedures \cite{Zienkiewicz1987} and also confirmed that the gradients of the thickness deformations 
induced by MscL lattices are sufficiently small for the standard leading-order model in equation~(\ref{eq:elastic_energy}) 
to be valid (see Supplementary Information Sec.~S1 for further details).
In the special cases for which analytic results on the minima of equation~(\ref{eq:elastic_energy}) are available 
\cite{Huang1986,CAH2013a}, our finite element procedure yields excellent agreement with exact analytic solutions.

\subsection*{Simulated annealing Monte Carlo simulations}
\label{secMC}
To confirm our predictions of the minimum-energy lattice architectures of tetrameric and pentameric MscL we carried out Monte Carlo simulations with simulated annealing of pair interaction
potentials \cite{Frenkel2001,Press2007}. 
To efficiently implement the simulations, we first used our finite element approach to calculate the thickness-mediated pair interaction energies $G_{\text{int}}(d, \omega_1, \omega_2)$ between closed and open tetrameric and pentameric MscL, where $d$ is the center-to-center distance between the two MscL and $\omega_{1,2}$ are the MscL orientations.
We used a translational resolution $\Delta d = 0.25$~nm and an orientational resolution $\Delta \omega = 3^\circ$, from which we constructed an array of interaction energies. We then approximated the interaction energy for arbitrary values of $(d,\omega_1, \omega_2)$ by first finding the appropriate $d$ row of the interaction energy array by rounding $\omega_1$ and $\omega_2$ to their closest calculated values, and then linearly interpolating the energy around $d$.
For fast evaluation of hard-core steric constraints, we constructed an analogous array for the minimum allowed distances $d_{\text{st}}(\omega_1, \omega_2)$.
Since, for the parameter values relevant for MscL, thickness-mediated interaction energies effectively vanish for $d\gtrsim15$~nm, we implemented cell list structures \cite{Frenkel2001} to accelerate pair evaluations.

In our simulated annealing Monte Carlo simulations, a single Monte Carlo step consists of one displacement and one rotation trial per MscL on average. We used a unit displacement $\delta d = 0.1$~nm and a unit rotation $\delta\theta=3^\circ$, for which about half of all Monte Carlo moves are accepted at $T=T_\text{rm}$, where $T_\text{rm}=298$~K is the room temperature.
The trials are accepted or rejected according to the Metropolis algorithm. In a typical run, we used $10^6$ Monte Carlo steps, and
decreased the temperature linearly starting from around $T=5 T_\text{rm}$ to $T=0$ during simulated annealing.
For minimization of pair interaction potentials with respect to only orientational degrees of freedom (see Supplementary Fig.~S7), 
we first initialized the system in the lattice symmetry of interest, and then set $\delta d = 0$ and only performed rotational Monte Carlo moves. We checked that all our results are robust with respect to different magnitudes of trial moves and different cooling schemes.

\section*{Pairwise additivity}
\begin{figure}
\center
\includegraphics[width=0.49\textwidth]{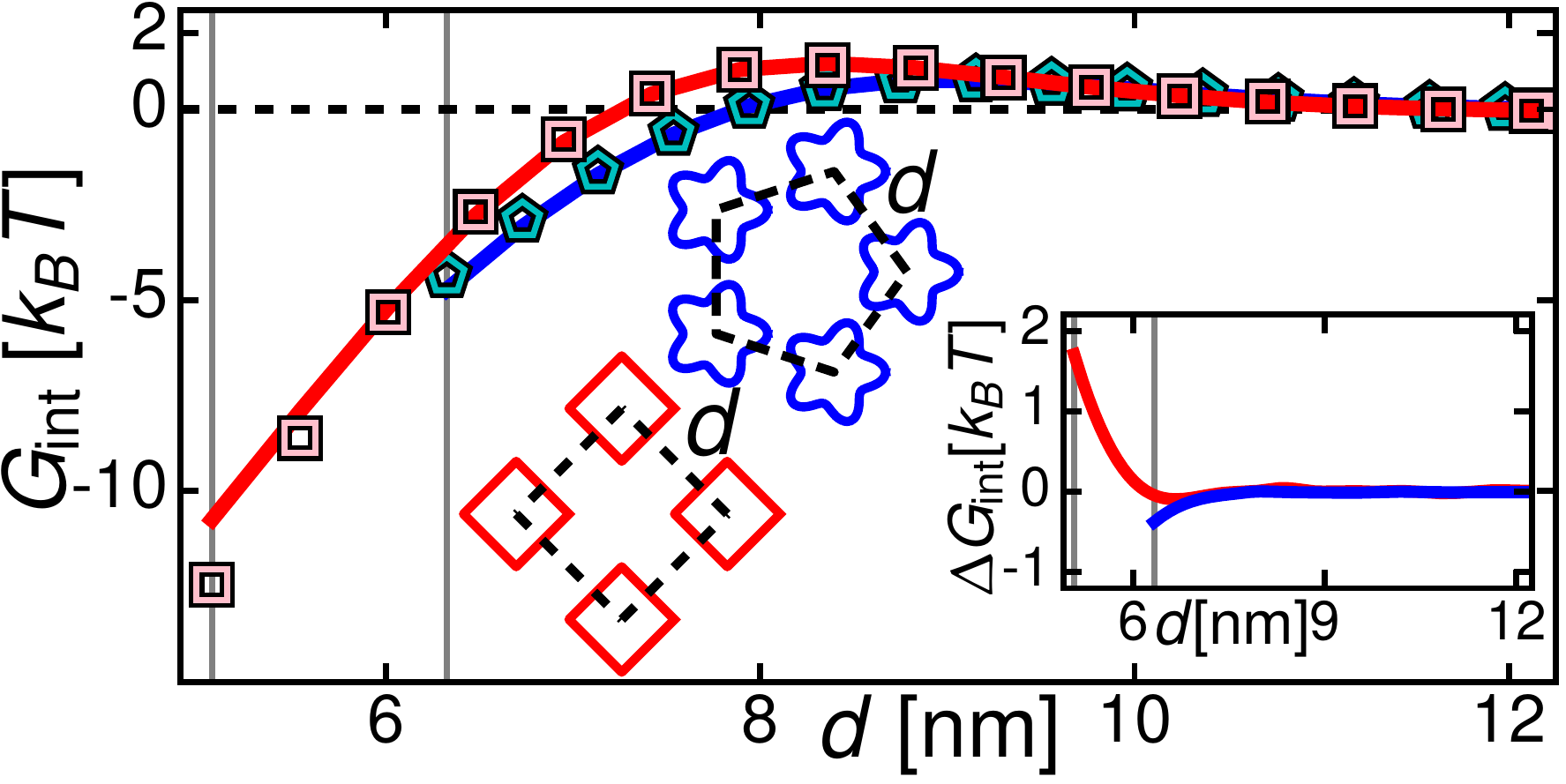}
\caption{\label{fig:pair} \textbf{Pairwise additivity of thickness-mediated protein interactions.} Thickness-mediated interaction energy, $G_\textrm{int}$, per closed MscL obtained from equation~(\ref{eq:elastic_energy}) for four tetrameric MscL and five pentameric MscL (solid curves),~and corresponding pairwise interaction energies (squares and pentagons), versus center-to-center distance between neighbouring MscL, $d$. Inset: Difference between multi-body and two-body interaction energies,~$\Delta G_\textrm{int}$, versus $d$. The vertical lines $d=d_\textrm{st}$ indicate steric constraints on MscL~configurations. We use the same MscL arrangements as in Fig.~\ref{fig:surfaces}.
}
\end{figure}

For curvature- and fluctuation-mediated interactions it has been suggested \cite{Kim1998,Dommersnes1999,Kim2008,Weitz2013,Yolcu2014} that non-pairwise contributions to the interaction energy can affect the stability of protein clusters. 
We find that non-pairwise contributions to thickness-mediated interactions modify the interaction strength but, except in special cases (see Supplementary Information Sec.~S3), do not alter how interactions vary with the shape and arrangement of proteins (see Fig.~\ref{fig:pair}).
Consistent with the corresponding two-body potentials \cite{Ursell2007,CAH2013a,OWC1}, the multi-body interactions between closed MscL in Fig.~\ref{fig:pair} are weakly unfavorable for center-to-center distances between neighbouring MscL, $d$, which are greater than $d\approx7.2$--$7.9$~nm (depending on MscL shape and orientation), and strongly favorable for smaller values of $d$. For fixed protein shape and orientation, thickness-mediated interactions are most favorable for the smallest value of $d$ allowed by steric constraints on lipid size, $d=d_\textrm{st}$, which corresponds to a minimum edge-to-edge protein separation of $\approx 1$~nm. At small~$d$, non-pairwise contributions to thickness-mediated interactions can be $>1k_B T$ in magnitude and, depending on protein shape and configuration, increase as well as decrease the interaction~energy (see Supplementary Information Secs.~S2--S4 for further details).

\section*{Lattices of tetrameric MscL}
Thickness-mediated MscL clustering was studied before\cite{Ursell2007, Grage2011} using the cylinder model of MscL\cite{Wiggins2004,Wiggins2005,Phillips2009}, which does not allow for the distinct symmetries of tetrameric and pentameric MscL observed in structural studies\cite{Liu2009, Chang1998, Walton2013}. For completeness, we summarize here, before turning to tetrameric MscL, the preferred lattice architectures associated with cylindrical MscL. 
In the cylinder model of MscL, MscL-induced lipid bilayer deformations are isotropic about individual MscL and, hence, thickness-mediated interactions between MscL depend on the separation but not on the orientation of MscL \cite{Ursell2007}. Allowing for planar clusters of interacting MscL, favorable MscL lattice architectures may be provided by lattices with honeycomb (three-fold), square (four-fold), or hexagonal (six-fold) symmetry. Calculating thickness-mediated interactions between many cylindrical MscL, we find that the honeyomb lattice is preferred at intermediate $d$, and the close-packed hexagonal lattice with $d=d_\textrm{st}$ provides the ground-state lattice architecture\cite{Grage2011} (see Supplementary Fig.~S5(a)).

\begin{figure}
\center
\includegraphics[width=0.49\textwidth]{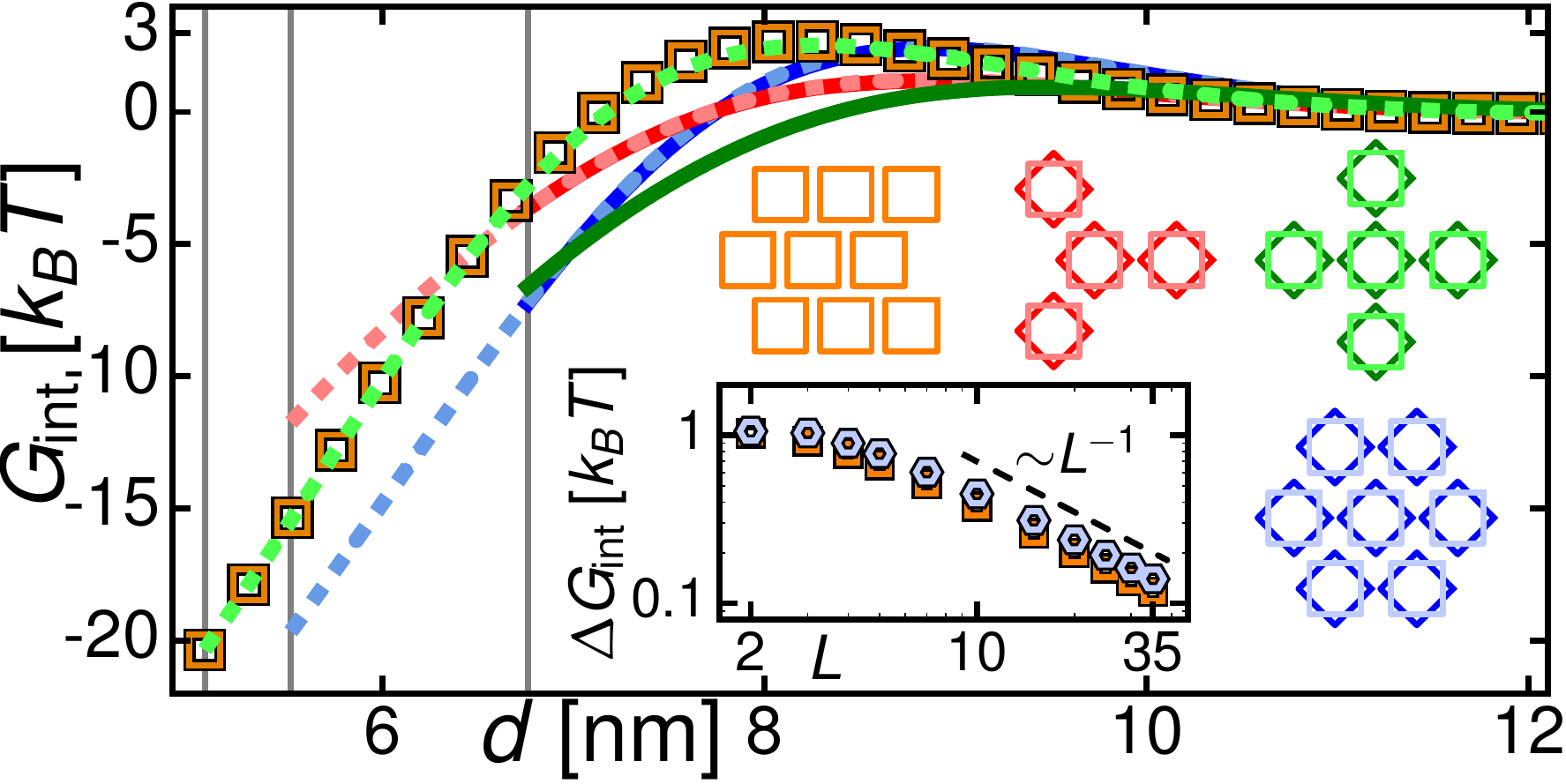} 
\caption{\textbf{Lattice architecture of tetrameric MscL.} Thickness-mediated interaction energy per closed
tetrameric MscL, $G_\textrm{int}$, versus center-to-center distance between neighbouring MscL in infinite honeycomb, square, and hexagonal lattices for face-on (dashed curves) and tip-on (solid curves) orientations of MscL, and in the shifted square lattice (squares). Honeycomb, square, and hexagonal lattices have three, four, and six nearest neighbours per lattice site, respectively.
Vertical lines indicate $d=d_\textrm{st}$. Inset: Difference between the ground-state energies of shifted square (squares), and face-on hexagonal (hexagons), lattices and the face-on square lattice per closed tetrameric MscL, $\Delta G_\textrm{int}$, as a function of square-root of number of MscL, $L$, offset by the energy difference at $L\to \infty$. Boundary effects decay as $1/L$.}
\label{fig:lattice4}
\end{figure}

For the observed shapes of MscL \cite{Liu2009,Chang1998}, thickness-mediated interactions between MscL not only depend on the separation but also on the orientation of MscL \cite{CAH2013a,OWC1} and, as a result, are inherently directional (see Fig.~\ref{fig:surfaces}). 
In particular, in the case of clusters of tetrameric MscL, we find that the distinct symmetry of tetrameric MscL, and resulting directionality of thickness-mediated interactions, yield a characteristic large-scale architecture of tetrameric MscL lattices which is different from the lattice architecture implied by the cylinder model of MscL (see Fig.~\ref{fig:lattice4}).
We first consider infinite honeycomb, square, and hexagonal lattices of tetrameric MscL, for which we evaluate the interaction energy per MscL by constructing unit cells with, by symmetry, zero slope of $u$ normal to their bilayer boundaries (see Supplementary Information Sec.~S3 for further details). We find that the honeycomb, square, hexagonal, and shifted square lattice architectures of tetrameric MscL yield qualitatively similar but, depending on the relative orientation of neighboring MscL, quantitatively distinct lattice energies.
In particular, the face-on square lattice with $d=d_\textrm{st}$ and the corresponding shifted square lattice, which are equally densely packed, have very similar bilayer deformation footprints and provide the ground-state lattice architectures for infinite lattices of tetrameric MscL. Restricting the minimum allowed $d$ to values $>d_\textrm{st}$, we predict that the hexagonal and tip-on square lattices become favorable as the minimum allowed $d$ is increased.

For finite clusters of tetrameric MscL, we have explicitly calculated ground-state lattice energies up to a cluster size of $\approx 1000$ MscL (Fig.~\ref{fig:lattice4} inset), which corresponds to the upper limit on MscL number observed \textit{in vivo} \cite{Bialecka2012}. Comparison of finite and infinite lattice energies shows that cluster size can strongly affect the magnitude of the lattice energy per MscL, but does not affect the competition between square, honeycomb, and hexagonal lattice architectures (see Supplementary Fig.~S6). However, tetrameric MscL are misaligned at the boundaries of the shifted square lattice, which increases the energy density of its cluster boundaries compared to the face-on square lattice. We predict that, due to this boundary effect, the face-on square lattice has a lower energy than the shifted square lattice (by $>4 k_B T$), and thus provides the ground-state lattice architecture for finite clusters of tetrameric MscL.

\section*{Lattices of pentameric MscL}
\begin{figure}
\center
\includegraphics[width=0.49\textwidth]{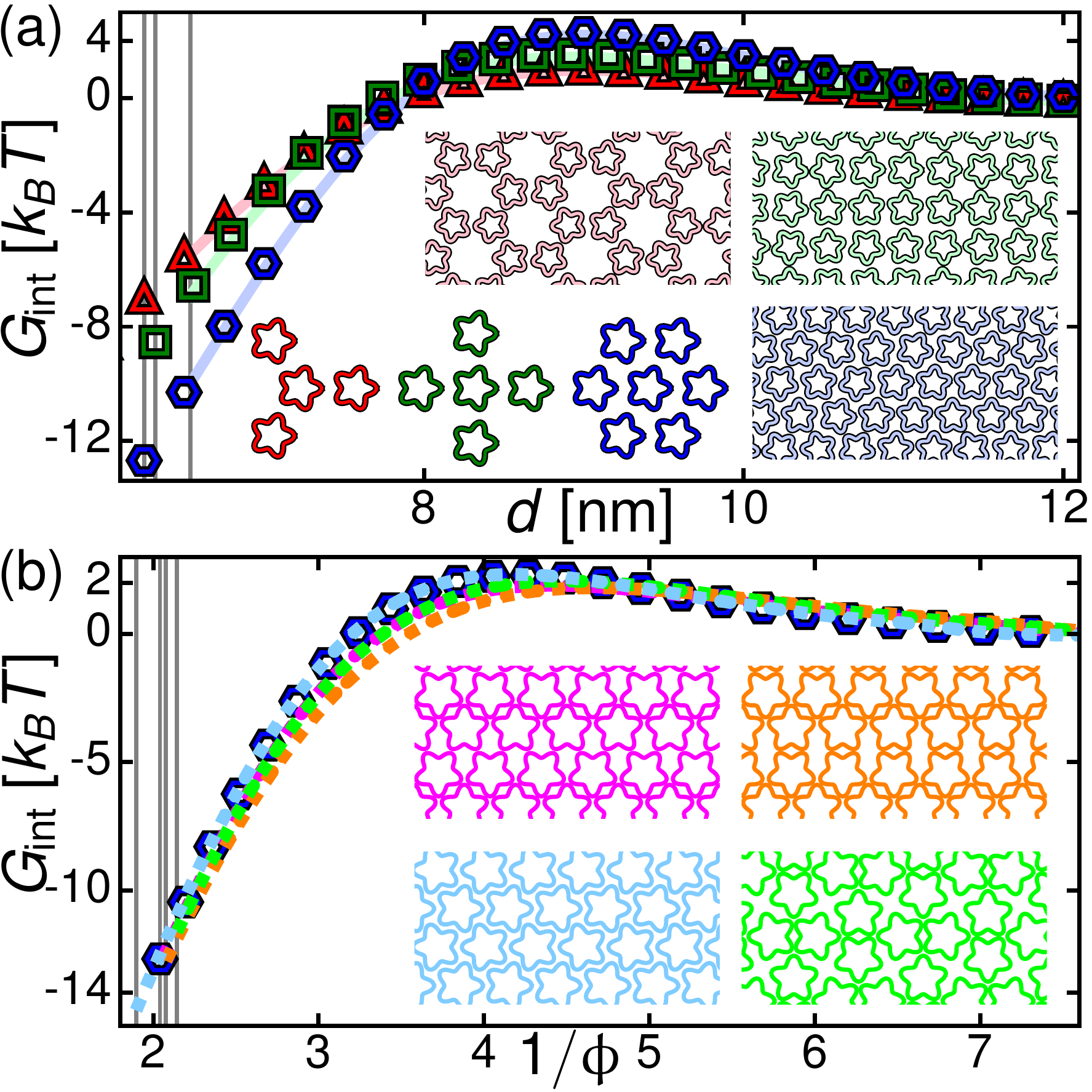}
\caption{\textbf{Lattice architecture of pentameric MscL.} Thickness-mediated interaction energy per closed
pentameric MscL, $G_\textrm{int}$, for (a) honeycomb, square, and hexagonal lattices versus center-to-center distance between neighbouring MscL, and (b) hexagonal and distorted \cite{Henley1986} lattices versus inverse area packing fraction. In (a), triangles, squares, and hexagons correspond to horizontally aligned MscL orientations (bottom left insets; $G_\textrm{int}$ for hexagonal
lattice reproduced in (b)) and solid curves to MscL orientations optimized at each $d$ through Monte Carlo simulations with simulated annealing of pair interaction potentials (top and right insets for a $d=d_\textrm{st}$ in the tip-on orientation of MscL). In (b), dashed curves correspond to the MscL packings shown in the insets. Based on the approximate cluster size observed \textit{in vitro} \cite{Grage2011}, we constructed pentameric MscL lattices from 216 MscL (honeycomb lattices), 220 MscL (bottom-right packing in (b)), and 225 MscL (all other cases) (see Supplementary Information Sec.~S4 for further details). Vertical lines indicate $d=d_\textrm{st}$.
}
\label{fig:lattice5}
\end{figure}

Pentameric MscL yield distinctively different lattice symmetries compared to tetrameric MscL (see Fig.~\ref{fig:lattice5}).
We first consider honeycomb, square, and hexagonal lattices with all MscL oriented in the horizontal direction, and with MscL orientations optimized at each $d$ by Monte Carlo simulations \cite{Frenkel2001} with simulated annealing \cite{Press2007} of pair interaction potentials (Fig.~\ref{fig:lattice5}(a)).
As in the case of thickness-mediated interactions between cylindrical MscL, the hexagonal (honeycomb) lattice is preferred at small (large) $d$ independent of the orientational ordering and cluster size considered (see Supplementary Information Sec.~S4). Thus, the directionality of thickness-mediated interactions between pentameric MscL \cite{CAH2013a,OWC1} does not affect the competition between honeycomb, square, and hexagonal lattice symmetries. Indeed, in planar lattices the five-fold symmetry of pentameric MscL necessarily leads to frustration of directional interactions \cite{Sachdev1985}.

However, allowing for distorted lattices with local orientational ordering \cite{Henley1986,Schilling2005,Atkinson2012} we find that the hexagonal lattice does not provide the ground-state lattice architecture for pentameric MscL (Fig.~\ref{fig:lattice5}(b)). Since the distorted lattices do not have a unique $d$ we compare lattice energies as a function of the area packing fraction $\phi$. Our results suggest that, at the largest $\phi$ allowed by steric constraints, an MscL arrangement similar to the closest packed of the distorted lattices (bottom-left lattice in Fig.~\ref{fig:lattice5}(b)), which corresponds to a distorted hexagonal lattice with alternate rows of MscL aligned in opposite directions, provides the ground-state lattice architecture, and that, in disperse clusters with variable $d$, face-on orientation of three neighbouring MscL (top-right lattice in Fig.~\ref{fig:lattice5}(b))
yields a favorable lattice architecture for a range of~$\phi$.

\section*{Simulated annealing of MscL clusters} 
\begin{figure}
\center
\includegraphics[width=0.49\textwidth]{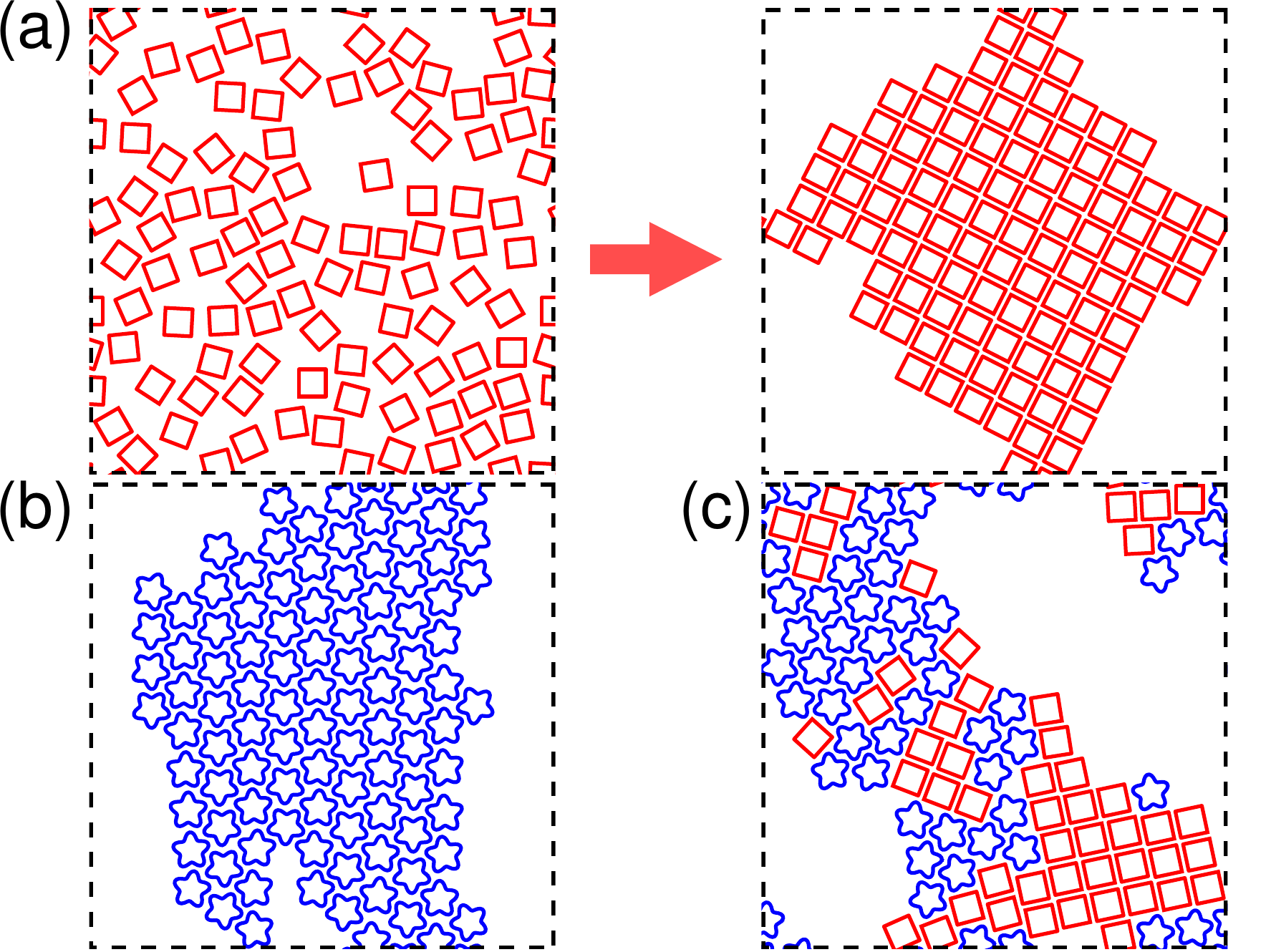}
 \caption{\textbf{Spontaneous ordering of MscL clusters through thickness-mediated interactions.} Ordering of closed (a) tetrameric (right panel and Supplementary Video S1), (b) pentameric (Supplementary Video S2), and (c) tetrameric and pentameric (Supplementary Video S3) MscL obtained through Monte Carlo simulations of translational and rotational diffusion with simulated annealing of pair interaction potentials. The left panel in (a) shows a typical (disordered) configuration used to initialize the simulations. We used periodic boundary conditions with 100 MscL.
 (See Supplementary Information Sec.~S5 for further details.)
}
\label{fig:annealing}
\end{figure}

We have confirmed our predictions of the minimum-energy MscL lattice architectures at $d=d_\textrm{st}$, and larger (fixed)~$d$, through Monte Carlo simulations \cite{Frenkel2001} of translational and rotational diffusion of MscL with simulated annealing \cite{Press2007} of pair potentials (see Fig.~\ref{fig:annealing}). In agreement with the multi-body calculations in Figs.~\ref{fig:lattice4} and~\ref{fig:lattice5} we obtain, in the ground state, face-on square lattices of tetrameric MscL (Fig.~\ref{fig:annealing}(a) and Supplementary Video S1) and distorted hexagonal ordering of pentameric MscL with alternate rows of MscL aligned in opposite directions (Fig.~\ref{fig:annealing}(b) and Supplementary Video S2). Subunit-counting experiments have suggested \cite{Gandhi2011,Walton2015} that, at least \textit{in vitro}, MscL can occur as a mixture of different oligomeric states. Simulated annealing of mixtures of tetrameric and pentameric MscL indicates that, in the ground state, tetrameric MscL forms a face-on square lattice in mixed MscL clusters (Fig.~\ref{fig:annealing}(c) and Supplementary Video S3), with the preferred distorted hexagonal arrangement of pentameric MscL being further distorted to accommodate tetrameric MscL lattices. These results also follow from Figs.~\ref{fig:lattice4} and~\ref{fig:lattice5} by noting that the ground-state lattice energy is lower for tetrameric than~pentameric~MscL.

\section*{Activation of MscL lattices}
\begin{figure}
\center
\includegraphics[width=0.49\textwidth]{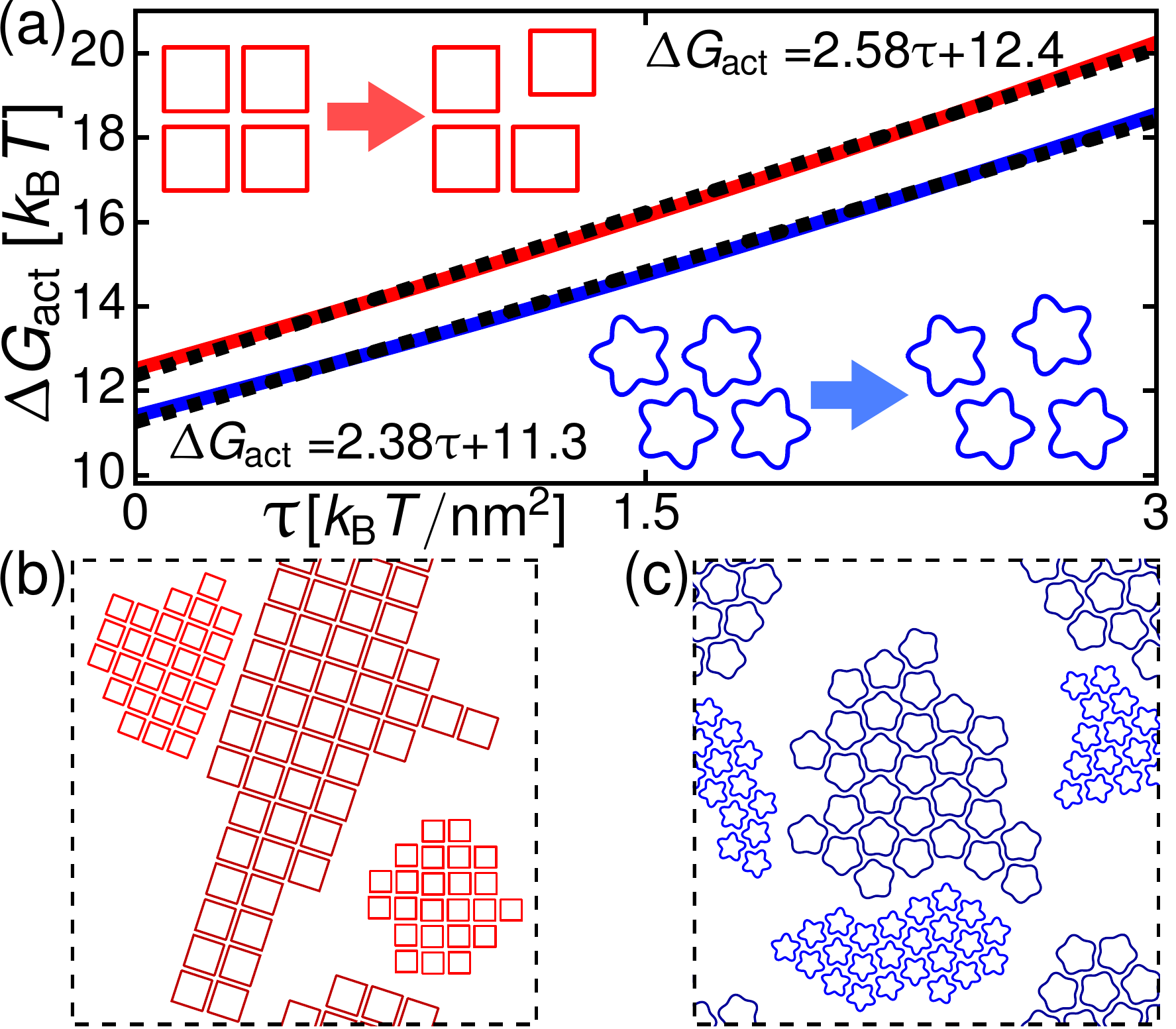}
\caption{\textbf{Gating of MscL lattices.} (a) Activation barrier, $\Delta G_\textrm{act}$, of tetrameric and pentameric MscL lattices versus membrane tension for the preferred lattice reorganizations suggested by the respective ground-state lattice architectures (insets). Dashed lines correspond to the indicated linear fits. (b,c) Architectures of tetrameric (Supplementary Video S4) and pentameric (Supplementary Video S5) MscL lattices of closed (smaller inclusions) and open (larger inclusions) MscL obtained by simulated annealing of pair interaction potentials. We used periodic boundary conditions with 50 closed and 50 open MscL.
}
 \label{fig:activation}
\end{figure}

Based on the available structural models of MscL in the open state, equation~(\ref{eq:elastic_energy}) predicts \cite{Wiggins2004,CAH2013b} a substantial difference in thickness deformation energy between the open and closed states of MscL, which accounts for the basic experimental phenomenology of MscL gating at dilute MscL concentrations \cite{Perozo2002,Phillips2009}. In crowded membranes, the gating of MscL clusters has been observed \cite{Grage2011} to be inhibited by an activation barrier, which slows the gating of MscL clusters. This activation barrier was attributed \cite{Grage2011} to the steric confinement of closed MscL in densely-packed MscL clusters. In particular, gating of MscL is accompanied by a substantial increase in the MscL cross-sectional area, by $\approx 20$~nm$^2$ per MscL \cite{Chiang2004}. As a result, when a cluster of closed MscL, assembled at small (or zero) membrane tension, is subjected to a large membrane tension, opening of MscL in the interior of the cluster would require an energetically costly large-scale reorganization of the lattice architecture to accommodate the increased cross-sectional area of open MscL. In contrast, for MscL lying along the cluster perimeter, only a relatively minor lattice arrangement is required to accommodate the open state (see Fig.~\ref{fig:activation}(a) insets), and the resulting activation barrier is small compared to MscL located in the cluster interior. We find that the magnitude of this activation barrier of ground-state lattices depends on MscL symmetry, and increases approximately linearly with membrane tension (see Fig.~\ref{fig:activation}(a)). 
The increase in the activation barrier of MscL lattices with increasing membrane tension in Fig.~\ref{fig:activation}(a) can be understood from an intuitive perspective by noting that an increase in membrane tension yields \cite{Ursell2007} a decrease in the preferred hydrophobic thickness of the lipid bilayer. The typical lipid bilayer considered here has a smaller preferred hydrophobic thickness than closed MscL, and the magnitude of the bilayer-MscL hydrophobic mismatch therefore increases with increasing membrane tension. This results in an increased magnitude of favorable interactions between closed MscL \cite{Ursell2007}, and a corresponding increase in the activation barrier of MscL lattices with increasing membrane tension.
The activation barrier is lowest at the corners of MscL lattices, and is higher (by $\approx 2\,k_B T$) for tetrameric than pentameric MscL lattices. Assuming an Arrhenius form for the reorganization rate of MscL lattices, these results imply that activation of tetrameric MscL lattices is slower by approximately one order of magnitude than activation of pentameric MscL lattices.

Structural models of MscL gating suggest \cite{Liu2009,Walton2015} that closed and open MscL have distinct hydrophobic thicknesses, yielding \cite{Ursell2007,CAH2013a,OWC1} 
weakly favorable thickness-mediated interactions at intermediate $d$, and strongly unfavorable interactions at small $d$. To study the ground-state
lattice architectures of partially activated MscL clusters \cite{Grage2011} we extended our simulated annealing simulations of translational
and rotational diffusion to include open as well as closed MscL. We find that, in agreement with experimental observations and previous calculations \cite{Grage2011}, closed and open MscL form composite clusters, but segregate into distinct sub-clusters (see Fig.~\ref{fig:activation}(b,c) and Supplementary Videos S4 and S5). Based on existing models of the shape of open MscL \cite{CAH2013b,Walton2015,Liu2009}, our simulations suggest that each sub-cluster of closed or open MscL shows the ground-state lattice architecture of tetrameric or pentameric MscL in
Figs.~\ref{fig:lattice4} and~\ref{fig:lattice5}, and that neighbouring closed and open MscL are separated by a characteristic center-to-center distance $d\approx9.5$~nm (see Supplementary Information Sec.~S7 for further details). 

\section*{Summary and conclusions}
We find that bilayer-mediated elastic interactions can yield ordering of mechanosensitive membrane protein clusters, linking membrane protein shape to the architecture and collective function of membrane protein lattices. We followed here experiments on MscL clustering \cite{Phillips2009,Grage2011,Nomura2012} and focused on minimum-energy lattice architectures due to thickness-mediated interactions between MscL \cite{Harroun99,Goforth2003,Botelho2006,Dan1993,ArandaEspinoza1996,harroun99b,Fournier1999,Partenskii2004,Brannigan2007,Ursell2007,CAH2014a}. In general, thermal fluctuations, membrane heterogeneity, and molecular effects not captured by the continuum approach \cite{Parton2011,Periole2007,Mondal2013,Yoo2013,West2009,Kim2012}, as well as curvature- \cite{Goulian1993,Weikl1998,Kim1998,Kim2008,Auth2009,Muller2010,Frese2008,Reynwar2011,Fournier1999,Dommersnes1999,Evans2003,Bahrami2014,Weitz2013,Yolcu2014}
and fluctuation-mediated \cite{Dommersnes1999,Evans2003,Weitz2013,Yolcu2014,Golestanian1996,Weikl2001,Lin2011} 
interactions, may also affect the architecture and function of membrane protein lattices. In particular, due to the frustration of directional interactions in pentameric MscL lattices, the local orientational ordering of pentameric MscL in the predicted distorted hexagonal lattices may be perturbed substantially by thermal fluctuations. More generally, thermal fluctuations will diminish long-range order in MscL lattices, and hence the predicted MscL lattice architectures will only be preserved locally (see Supplementary Information Sec.~S6). 
Previous theoretical estimates suggest \cite{Ursell2008} that fluctuation-mediated interactions between MscL, while weak compared to thickness-mediated interactions, are favorable, and thus might further stabilize MscL clusters.

We predict that, for MscL clustering driven by thickness-mediated interactions \cite{Harroun99,Goforth2003,Botelho2006,Dan1993,ArandaEspinoza1996,harroun99b,Fournier1999,Partenskii2004,Brannigan2007,Ursell2007,CAH2014a}, tetrameric \cite{Liu2009} and pentameric \cite{Chang1998} MscL yield distinct lattice architectures and lattice activation barriers. In particular, our calculations suggest that, locally, clusters of tetrameric MscL show a four-fold symmetric translational ordering with neighboring MscL in a face-on orientation, while clusters of pentameric MscL show an approximately six-fold symmetric translational ordering with alternate rows of pentameric MscL aligned in opposite directions. We predict that, in mixed clusters of tetrameric and pentameric MscL, the preferred distorted hexagonal arrangement of pentameric MscL is further distorted to accommodate face-on square lattices of tetrameric MscL. Furthermore, we find that lattices of tetrameric MscL have a higher activation barrier than lattices of pentameric MscL and that, in both cases, the lattice activation barrier increases approximately linearly with membrane tension. Our calculations suggest that activation of tetrameric MscL lattices is slower by approximately one order of magnitude than activation of pentameric MscL lattices. 
Finally, we predict that MscL can form mixed clusters of closed and open MscL, with open and closed MscL segregated into distinct sub-clusters which show the face-on square or distorted hexagonal lattice architectures associated with tetrameric or pentameric MscL, respectively.
The predicted lattice architectures of mixed clusters of closed and open MscL may be experimentally accessible \textit{in vitro} through suitable modifications of bilayer-MscL interactions \cite{Grage2011,Perozo2002Nature,Perozo2002,Phillips2009}, but may not be accessible \textit{in vivo} due to the short lifetimes of MscL in the open state.

Our predictions may be most straightforward to test experimentally by extending existing \textit{in vitro} assays \cite{Grage2011,Nomura2012} for investigating MscL clustering to account for different oligomeric states of MscL. In particular, tetrameric and pentameric MscL, as well as mixtures of tetrameric and pentameric MscL, can be produced \textit{in vitro} \cite{Dorwart2010,Iscla2011,Gandhi2011,Reading2015} by selectively truncating MscL, tuning the lipid or detergent compositions, or varying the temperature. While not all MscL oligomeric states thus produced may be fully functional, such \textit{in vitro} studies may nevertheless allow direct experimental tests of the predicted relations between MscL symmetry and MscL lattice architecture. More speculatively, the relations between MscL oligomeric state, MscL lattice architecture, and MscL lattice activation barrier found here may provide novel approaches for probing the oligomeric state of MscL \textit{in vivo} to address whether MscL only occurs in its pentameric state \textit{in vivo} \cite{Dorwart2010,Iscla2011}, or whether MscL may occur as a mixture of different oligomeric states \textit{in vivo} \cite{Gandhi2011,Walton2015}. Considering that a wide range of membrane proteins are mechanosensitive \cite{Phillips2009,Anishkin2014,Brohawn2014,Kung2010}, we suggest that experiments on the link between MscL symmetry, lattice architecture, and collective lattice function predicted here will yield general insights into how membrane organization broadens the repertoire of protein function.


\section*{Acknowledgements}
We thank R. Phillips and D. C. Rees for helpful comments.
This work was supported at USC by NSF award number DMR-1206332, an Alfred P. Sloan Research Fellowship in Physics (to C.A.H.), and by the USC Center for High-Performance Computing, and at UCLA by NSF award numbers CMMI-0748034 and DMR-1309423.

\section*{Author contributions}
O.K., W.S.K., and C.A.H. designed the study, O.K and P.D.K. performed the calculations, O.K. and C.A.H. analysed the results.
All authors contributed to writing the manuscript.

\section*{Additional information}
Supplementary information accompanies this paper at \href{http://www.nature.com/articles/srep19214}{www.nature.com/articles/srep19214}.

\paragraph{Competing financial interests:} The authors declare no competing financial interests.

\end{document}